\documentclass[%
 reprint,
 amsmath,amssymb,
 prl,
]{revtex4-1}

\usepackage{graphicx}
\usepackage{dcolumn}
\usepackage{bm}

\newcommand{\units}[1]{\,\mathrm{#1}}

\newcommand{\JILA}{JILA, National Institute of Standards and Technology and the University of Colorado, Boulder, Colorado 80309, USA}
\newcommand{\NIST}{Department of Physics, University of Colorado, Boulder, Colorado 80309, USA}

\begin{document}


\title{Cavity quantum acoustic device in the multimode strong coupling regime}

\author{Bradley A. Moores }
\thanks{These two authors contributed equally}
\email{bradley.moores@colorado.edu}
\affiliation{\JILA}
\affiliation{\NIST}

\author{Lucas R. Sletten}%
\thanks{These two authors contributed equally}
\affiliation{\JILA}
\affiliation{\NIST}

\author{Jeremie J. Viennot}%
\affiliation{\JILA}
\affiliation{\NIST}

\author{K. W. Lehnert}%
\affiliation{\JILA}
\affiliation{\NIST}


\date{\today}

\begin{abstract}
We investigate an acoustical analog of circuit quantum electrodynamics
that facilitates compact high-Q (${>}20,000$) microwave-frequency cavities with dense spectra. We fabricate and characterize a device that comprises a flux tunable transmon coupled to a $300 \units{\mu m}$ long surface acoustic wave resonator. 
For some modes, the qubit-cavity coupling reaches $\units{6.5\units{MHz}}$, exceeding the cavity loss rate ($200\units{kHz}$), qubit linewidth ($1.1\units{MHz}$), and the cavity free spectral range ($4.8\units{MHz}$), placing the device in both the strong coupling and strong multimode regimes.
With the qubit detuned from the cavity, we show that the dispersive shift behaves according to predictions from a generalized Jaynes-Cummings Hamiltonian. 
Finally, we observe that the qubit linewidth strongly depends on its frequency, as expected for spontaneous emission of phonons, and we identify operating frequencies where this emission rate is suppressed.

\end{abstract}

\pacs{Valid PACS appear here}
\maketitle



Emergent phenomena of many-body spin physics may be studied more readily with artificial systems rather than real materials. This possibility has led to a proliferation of techniques striving to emulate model Hamiltonians that exhibit many-body localization \cite{Dalichaouch1991, Wiersma1997, Lye2005, Fort2005}, topological protection \cite{Lu2009, Rechtsman2013}, and quantum phase transitions \cite{Greiner2002,Roushan2017, Fitzpatrick2017}.
Of these artificial systems, transmon qubits coupled by microwave frequency electrical resonances show tremendous promise \cite{Houck2012}. Specifically, if microwave excitations of the qubits or resonators are regarded as particles, the strong electrical non-linearity of the qubit creates an effective particle-particle interaction that is much larger than the relevant dissipation and decoherence processes \cite{Schmidt2013}. 
Such an equivalently strong interaction has yet to be demonstrated with optical photons. Furthermore, in contrast to artificial systems that hold single atoms in optical lattices \cite{Goban2015}, planar circuits are rigidly fixed to their substrate and therefore have no spatial entropy.

Coupling many qubits to a dense cavity spectrum has been proposed as a means of engineering finite-range interactions for use in analog quantum simulations \cite{Strack2011}.
However, building such a system in the circuit quantum electrodynamics (cQED) architecture \cite{Wallraff2004, Blais2004, McKay2015} is hindered by the mismatch of scales between the qubits and the electromagnetic modes.
For example, low dissipation planar resonators are centimeter long transmission lines \cite{Frunzio2005}, whereas the transmons are generally ${\sim}100 \units{\mu m}$. Furthermore these planar resonators are difficult to shield from each other, often resulting in undesired couplings.

As sound propagates 5 orders of magnitude slower than the speed of light, this scale mismatch can be overcome by replacing electromagnetic resonators with acoustic cavities, a strategy that has been pursued with bulk acoustic waves \cite{Connell2010,Chu2017}.
Surface acoustic waves (SAWs) \cite{Campbell1989} have the additional feature that they are
confined to the surface of a chip, allowing them to interact with sophisticated planar structures and many qubits. They readily make compact, multimode cavities with excellent shielding.
At low temperatures and with excitations on the single phonon scale, SAW cavities have been demonstrated with high quality factors \cite{Magnusson2015,Manenti2015}.
Transmon qubits have been successfully coupled to propagating SAWs on GaAs \cite{Gustafsson2014}, and to a single mode of a SAW resonator on quartz \cite{Manenti2017, Noguchi2017, Bolgar2017}, but presently it's unclear what limits the coherence of acoustically coupled qubits.
In order for qubits in cavity quantum acoustodynamical (CQAD) systems to experience coherent and finite range interactions, the system should operate in both the strong coupling and strong multimode limits.

\begin{figure*}[htb]
\begin{center}
\includegraphics[width=1.0\linewidth]{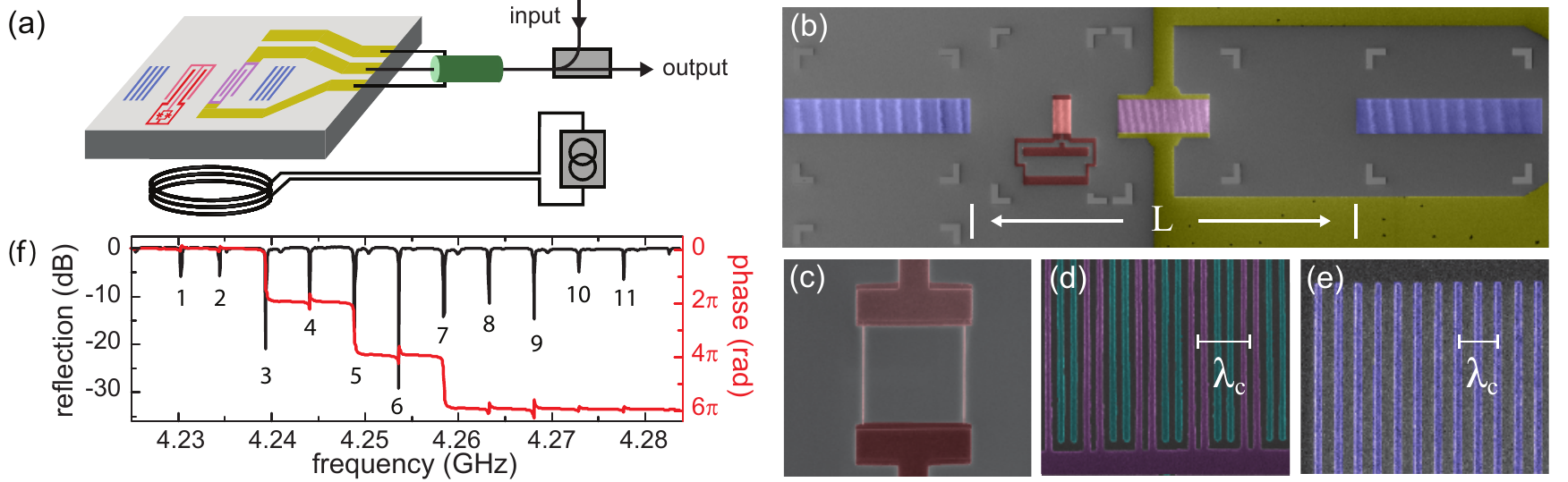}
\caption {\small 
Device diagram and acoustical cavity spectrum. (a) A cartoon schematic of the SAW cavity, acoustically coupled qubit, and the microwave network for control and measurement. 
(b) A false color SEM image of the fabricated device before the Josephson junction was patterned.
Two Bragg reflectors (blue) are spaced by $L=275\units{\mu m}$ to define a SAW cavity, and a split-junction transmon qubit (red) was placed at $L_\mathrm{eff}/4$ from the left reflector. A cavity-IDT (pink) at $L/2$ is used to drive and readout the cavity modes. The center- and ground-conductor of a coplanar waveguide (yellow) contact either side of the IDT.
Measurements consist of detecting the reflection of a microwave tone applied to the cavity-IDT. A directional coupler separates incident and reflected waves, so that the reflected signal is passed through a high electron mobility transistor (HEMT) amplifier and measured. 
False color SEM images of (c) a split Josephson-junction with a $7{\times}7\units{\mu m^2}$ loop area, (d) a split-finger IDT with the upper electrode in green and the lower in purple \cite{supp}, and (e) several Al stripes within a Bragg reflector.
The characteristic wavelength of the cavity is indicated, $\lambda_c = v_s / f_c = 677\units{nm}$, where the center frequency is $f_c = 4.253\units{GHz}$.
(f) A microwave reflection measurement of the SAW cavity reveals 11 (numbered) prominent longitudinal modes within the mirror bandwidth.}
\label{fig1}
\end{center}
\end{figure*}

In this Letter, we demonstrate such a CQAD system where the qubit-cavity coupling strength is larger than both the qubit and cavity decoherence rates, and larger than the cavity free spectral range (FSR) $\nu_\mathrm{FSR}$.  Furthermore, we show we can overcome an essential incompatibility between long qubit coherence and operation in the dispersive limit of a strong multimode system, where multiple qubits could emulate exchange coupled spins \cite{Blais2007}.  In detail, when the qubit is resonant with the spectrum of acoustic modes, we observe clear avoided crossings and extract the couplings $g_m$ of the qubit to $17$ high quality modes of the acoustic cavity, finding $g_m/2\pi \sim \nu_\mathrm{FSR}$ for most modes.  This strong multimode coupling inhibits the qubit from reaching the dispersive regime when its frequency lies between these modes. However, the cavity confines phonons only over a narrow frequency range, allowing the qubit to be far detuned from all resonant modes while also relaxing the qubit via phonon radiation. Indeed, in the dispersive regime we measure the qubit linewidth as a function of qubit frequency and resolve a substantial contribution from spontaneous emission of unconfined phonons \cite{Kockum2014}. But crucial to the feasibility of  many-body spin emulation, we also identify special qubit frequencies where this emission is prohibited.

We demonstrate these characteristics with the device drawn schematically in Fig.\ \ref{fig1}(a) and imaged in Fig.\ \ref{fig1}(b). This device is a flux tunable qubit inside a multimode SAW cavity on GaAs. 
The qubit is a transmon consisting of a split Josephson junction in parallel with a split-finger interdigitated transducer (IDT). The IDT forms both a shunting capacitor (${\sim}100\units{fF}$) and a piezoelectric transducer that interacts with SAW waves. 
The cavity is defined by two Bragg reflectors separated by $275\units{\mu m}$, each consisting of a periodic array of aluminum stripes. Each stripe weakly reflects incoming SAWs (${<}2\%$), primarily due to mass loading. 
The arrays are highly reflective over a narrow frequency range (${\sim}50 \units{MHz}$), while SAW penetration makes the effective cavity length $L_\mathrm{eff}= 300 \units{\mu m}$.
The acoustic response is probed through a split-finger IDT, centered in the cavity, that converts between mechanical excitations in the cavity and microwave signals in the coplanar waveguide.

We first characterize the acoustic modes by tuning the transmon far away from the cavity resonances using an off-chip coil. The device was embedded in a microwave measurement network as shown in Fig.\ \ref{fig1}(a) and cooled below $30\units{mK}$ in a dilution refrigerator.
Fig.\ \ref{fig1}(f) shows the microwave reflection coefficient versus frequency of the acoustic cavity. Over the mirror bandwidth of approximately $50 \units{MHz}$, we observe 11 prominent equally spaced resonances. For each of these dips, there are weaker adjacent resonances at higher frequency. We interpret the 11 prominent resonances as purely longitudinal cavity modes, and the higher frequency satellites as modes with a non-zero transverse mode number. 
In what follows, we will model the 11 longitudinal modes and the 6 more visible transverse modes. From the spacing between longitudinal modes we extract the cavity FSR $\nu_\mathrm{FSR}=v_s/2L_{\mathrm{eff}}=4.8\units{MHz}$, where $v_s=2880\units{m/s}$ is the speed of sound on GaAs, consistent with our expectation from the cavity geometry. The longitudinal modes have $\kappa_l/2\pi \approx 200\units{kHz}$ linewidths, and the transverse mode linewidths are slightly lossier with  $\kappa_t/2\pi \approx 400\units{kHz}$.


\begin{figure*}[htb]
\begin{center}
\includegraphics[width=1.0\linewidth]{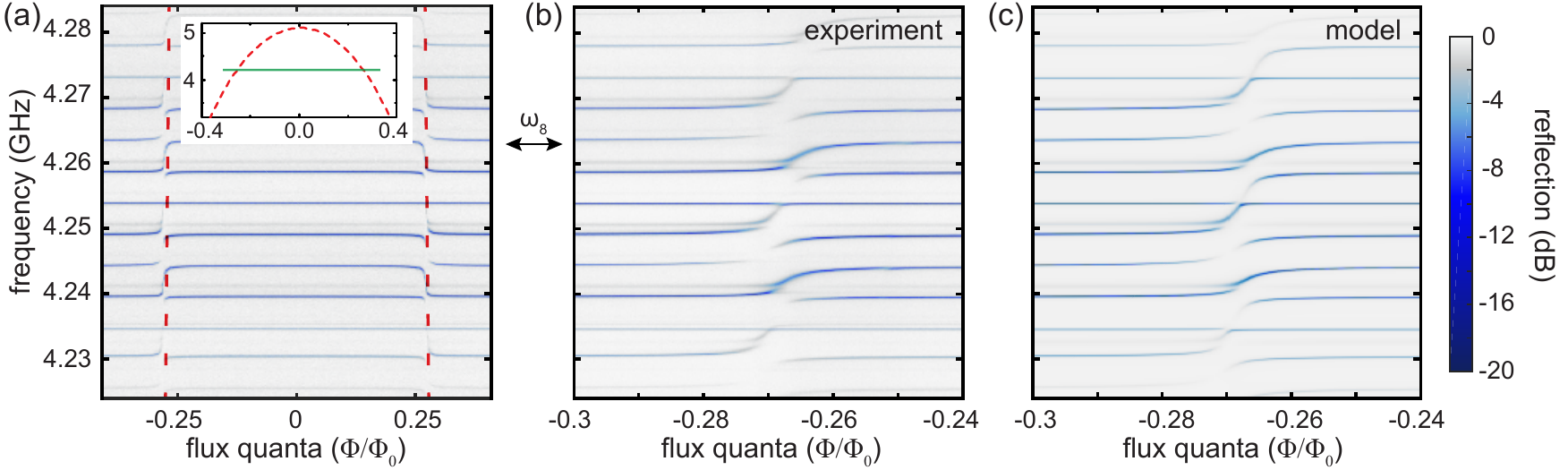}
\caption {\small Resonantly coupled multimode CQAD system. (a) The plot shows the cavity reflection (color scale) as the transmon is tuned by varying the applied magnetic flux. 
The red dashed line indicates the transmon's flux dependent resonance. The inset shows the expected qubit frequency (red), which has a maximum at $5.08\units{GHz}$. The green section indicates the measured range of panel (a).
(b) Zoom in on the first set of avoided crossings shown in panel (a).
(c) A model for the acoustic spectrum based on the interaction Hamiltonian, Eq.\ (2). 
}
\label{fig2}
\end{center}
\end{figure*}


Having characterized the bare cavity spectrum, we tune the qubit into resonance with the modes to measure the transmon-cavity coupling strengths.
Figure \ref{fig2}(a) shows the cavity response as the coil current is swept, revealing two sets of avoided crossings at $\pm 0.27 \units{\Phi/\Phi_0}$. 
The qubit position in the cavity has significant consequences on the spectrum as the coupling strengths $g_m$ depend on the spatial overlap of a mode and the qubit-IDT fingers.
For example, the transmon couples strongest to modes that have anti-nodes aligned 
with the qubit-IDT fingers.  
Zooming into a single set of avoided crossings  (Fig.\ \ref{fig2}b) reveals that the transmon indeed couples to the cavity modes with varying strength.


Although the spectrum looks complicated, the longitudinal mode couplings exhibit a simple oscillating pattern. For example, modes 4 and 8 strongly couple, modes 2, 6, and 10 barely couple, and the odd modes all moderately couple. 
The qubit position at $L_\mathrm{eff}/4$ creates a mode-dependent coupling strength
\begin{equation}
    g_m = g_0 \sin\left (\frac{\pi}{4}m+\phi_q\right)
\label{coupling}
\end{equation}
where $g_0$ is an overall coupling strength that is sinusoidally modulated by a four mode period, and $\phi_q$ is an overall phase shift set by the small deviation in the qubit position from $L_\mathrm{eff}/4$. 
Coupling to the transverse modes can be written in a similar way, with the same phase $\phi_q$ and a smaller $g_0$ which can be approximated from the cavity-IDT spectrum \cite{supp}.

Using this insight, we make a simple model of the multimode cavity and qubit system. It can be described by an $18{\times}18$ interaction Hamiltonian
%
%
\begin{equation}
 H/\hbar = \begin{pmatrix} \omega_1 & & & g_1\\  
    & \omega_2 & & g_2\\  
    & & \ddots  & \vdots \\ 
    g_1 & g_2 & \cdots & \omega_{\mathrm{q}}
    \end{pmatrix} 
\label{hamiltonian}
\end{equation}
where $\omega_k/2\pi$ are the 17 uncoupled cavity modes (11 longitudinal and 6 transverse), and $\omega_{\mathrm{q}}/2\pi$ is the qubit ground to first excited state transition frequency.
The number of coupling terms, and consequently fit parameters, can be significantly reduced from 17 to 3 using Eq.\ (\ref{coupling}) and an equivalent equation for the transverse modes. We diagonalize the Hamiltonian as a function of the qubit frequency to obtain the hybridized modes.
We found the optimum fit is $g_0/2\pi= 6.5\units{MHz}$ and $\phi_q ={-}0.1\units{rad}$, which is plotted in Fig.\ \ref{fig2}(c). The model indicates that modes 4 and 8 couple strongest to the transmon with $g_{4,8}/2\pi=6.48\units{MHz}$. Because the coupling strength of some modes exceed the cavity FSR ($\nu_{FSR} =4.8\units{MHz}$), the device seems to operate in the strong multimode regime.

\begin{figure}[htb]
\begin{center}
\includegraphics[width=1.0\linewidth]{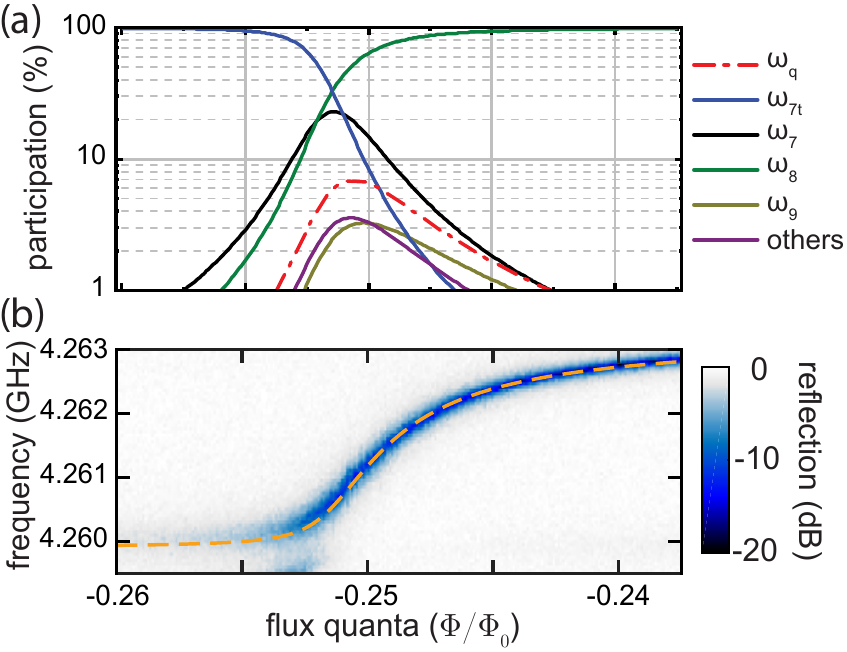}
\caption {\small 
Hybridization of modes in the strong multimode regime.
(a) The solid lines show the squared coefficients of the eigenvector (participation) in the uncoupled mode basis as the qubit tunes with magnetic field.
At $-0.252$ flux quanta, longitudinal modes $\omega_{7}$ and $\omega_{8}$ and transverse mode $\omega_{7t}$ nearly equally contribute to $89\%$ of the eigenvector. The `others' trace shows the combined contribution of the remaining modes. 
(b) The acoustic spectrum (from Fig.\ \ref{fig2}b) between modes $\omega_{7}$ and $\omega_{8}$. The dotted line shows the eigenvalue fit from the model corresponding to the superposition of modes from (a).
}
\label{fig3}
\end{center}
\end{figure}

The hallmark of the strong multimode limit consists of many modes hybridizing with each other through a mutual qubit coupling, 
while the qubit participation in each eigenmode remains low
\cite{Sundaresan2015}.
We can use our model Hamiltonian to infer that our device operates in this limit. 
In Fig.\ \ref{fig3}(a) we plot the qubit and acoustic mode participation in an eigenstate as the qubit frequency varies with magnetic field. 
On resonance, three modes strongly hybridize, where each mode almost equally contributes to the eigenvalue, while the qubit participation remains small (${<}7\%$).
Fig.\ \ref{fig3}(b) shows agreement between the data and model for the hybridization shown in Fig.\ \ref{fig3}(a).

Because the qubit participation is low, the well resolved avoided crossings do not imply that the CQAD system reaches the strong coupling limit ($g_0 > \{\kappa, \gamma\}$). To show that it does, we measure the qubit linewidth by operating the device in the dispersive limit \cite{Koch2007}. 
We begin by detuning the qubit far from all of the cavity resonances, which is possible because the mirrors that define the cavity are narrowband (Fig. 1f).  We then apply two tones to the cavity-IDT, one resonant with the eighth longitudinal mode of the cavity and one nearly resonant with the qubit. By monitoring the reflection of the tone at $\omega_8$, while varying the frequency and power of the qubit drive we detect the qubit's resonance through the qubit-state-dependent dispersive shift $\chi$ of the cavity resonance.


\begin{figure}[htb]
\begin{center}
\includegraphics[width=1.0\linewidth]{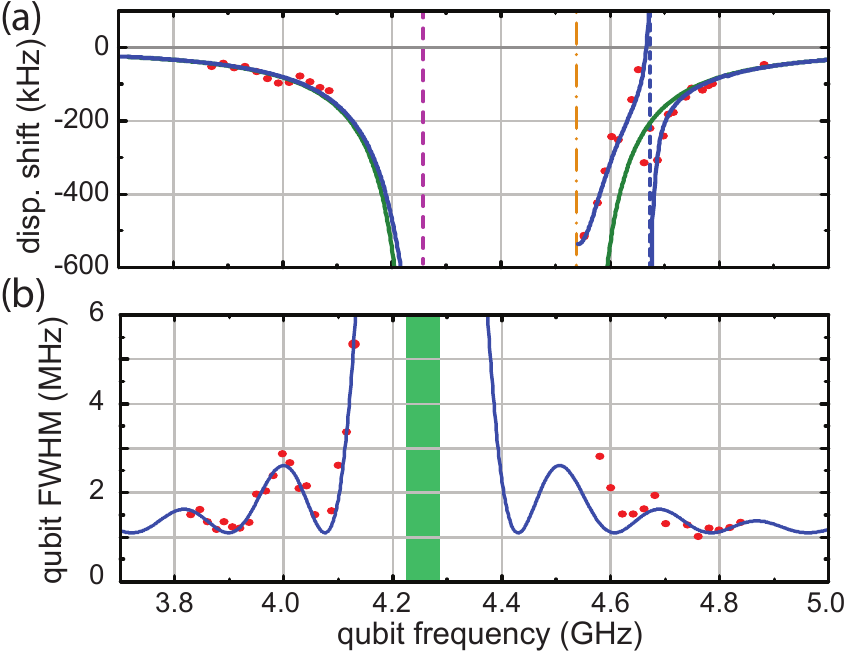}
\caption {\small 
Measurements in the dispersive regime. 
(a) The dispersive shift of cavity mode $\omega_8$ versus qubit frequency ($\omega_{\mathrm{q}}$). The purple, orange, and blue dashed lines indicate $\omega_\mathrm{8}/2\pi$, $\omega_\mathrm{8}/2\pi + \alpha$, and $\omega_\mathrm{8}/2\pi + 3\alpha/2$ respectively, where $\alpha=273\units{MHz}$ is the transmon's anharmonicity. 
The green line shows a prediction from the standard transmon model, and the solid blue line shows a prediction from diagonalizing a generalized Jaynes-Cummings Hamiltonian with a 4-level transmon.
(b) The qubit linewidth versus qubit frequency. Red points are measurements, and the blue line shows the expected spontaneous emission rate of the qubit into SAWs based on the qubit-IDT geometry, with an offset to account for intrinsic dephasing. The green region indicates the mirror bandwidth.
}
\label{fig4}
\end{center}
\end{figure}

Using this dispersive measurement of the qubit's state, we flux tune the qubit's resonance frequency and verify that our CQAD system behaves according to a generalized Jaynes-Cummings Hamiltonian. Specifically, at each value of applied flux, we measure the qubit frequency shift as function of cavity drive power (Stark shift).
In the low power limit the Stark shift is linear with phonon number, with a slope of $2\chi$ \cite{Schuster2005, supp}.
Figure \ref{fig4}(a) compares $\chi$ measurements of the CQAD device to predictions from two models.
When $\omega_{\mathrm{q}}<\omega_\mathrm{8}$, our transmon is well described by the standard transmon dispersive model \cite{Koch2007}, which only takes the lowest three energy levels into account and ignores two-phonon transitions. We use this regime to calibrate the single phonon power level.
%
When $\omega_{\mathrm{q}}>\omega_\mathrm{8}$ and a large phonon occupation is used, other transition frequencies and higher order effects become significant. To take these effects into account, we model the dispersive shift by diagonalizing a generalized Jaynes-Cummings Hamiltonian consisting of a 4-level transmon and a harmonic oscillator truncated at 50 excitations. 

With qubit spectroscopy well modeled by transmon theory, we can use the qubit linewidth measured in the low power limit of the cavity and qubit drives as an upper bound on the qubit decoherence rate. Unlike a system where a cavity fully encloses a qubit \cite{Paik2011}, the CQAD device interacts with unconfined modes outside of the mirror bandwidth that could limit the qubit coherence. 
However, the qubit transition can be tuned to specific frequencies in which the emission can be strongly suppressed.
The spatial periodicity and finite length of the qubit-IDT combine to emit SAWs with wavelengths centered around $\lambda_c$.
In the frequency domain, the IDT's spatial periodicity yields a SAW emission rate with a $\mathrm{sin}^2 X/X^2$ frequency dependence, where $X= N_q \pi (f{-}f_c)/f_c$ and $N_q$ is the number of qubit-IDT finger periods \cite{supp}. This spectrum consists of evenly spaced minima where emission is prohibited due to coherent cancellation.
The blue curve in Fig.\ \ref{fig4}(b) shows the expected qubit linewidth as a sum of the predicted spontaneous phonon emission rate of the transmon and a constant offset to account for intrinsic decoherence.
The red points show the inferred qubit linewidth, taken in the low-power limit. 
The qubit linewidth narrows when its frequency is within an IDT band minima (e.g.\ $3.9\units{GHz}$), and broadens by up to a factor of 3 near a maximum (e.g.\ $4.0\units{GHz}$).
From the offset between the predicted IDT spontaneous emission rate and the observed qubit linewidth, we estimate an upper-bound on the intrinsic qubit linewidth of $\gamma/2\pi = 1.1 \units{MHz}$.  Thus the qubit can exchange energy with a SAW mode at a rate about 6 times greater than its intrinsic decoherence rate, reaching the strong coupling limit.

In conclusion, we have shown that superconducting qubits and SAW cavities can reach the strong coupling and multimode regimes of CQAD while avoiding an incompatibility between qubit coherence times and dispersive operation. Although emulating many-body spin systems will require further improvement in qubit coherence, we show that decoherence from phonon emission can be strongly suppressed. In addition, this result may also allow coherent exchange of quantum states between atom-like defects or quantum dots and superconducting qubits, as proposed in some schemes to create a quantum electro-optical converter \cite{Schuetz2015}. Finally, with recent progress in improving the coherence of nano-mechanical resonators \cite{Meenehan2015}, these may supplant electromagnetic resonators in certain quantum information processing tasks, as they are much smaller and more easily isolated from one another when fabricated on the same chip.


\vspace{0.1in}
\noindent{\emph {Acknowledgment}} We gratefully acknowledge contributions from Xizheng Ma and Alex Martin. This material is based upon work supported by the National Science Foundation under Grant No. PHY 1734006.


%

\pagebreak
\widetext
\begin{center}
\textbf{\large Supplementary Material for\\ ``Cavity quantum acoustic device in the multimode strong coupling regime''}
\end{center}

\setcounter{equation}{0}
\setcounter{figure}{0}
\setcounter{table}{0}
\setcounter{page}{1}
\makeatletter
\renewcommand{\theequation}{S\arabic{equation}}
\renewcommand{\thefigure}{S\arabic{figure}}
\renewcommand{\bibnumfmt}[1]{[#1]}
\renewcommand{\citenumfont}[1]{#1}

\section{Fabrication}

The device was fabricated by patterning aluminum on (100) GaAs (AXT Inc.\,) using electron-beam (e-beam) lithography. A $6.2{\times}6.2\units{mm^2}$ chip was cleaved from a wafer, and the native oxide layer was removed by immersing it in ammonium hydroxide for $5\units{min}$. A $200\units{nm}$ PMMA layer (950PMMA A4 from MicroChem Corp.) was spin-coated at 4000 RPM and baked at $180\units{^{\circ}C}$ for $4\units{min}$. The acoustic layer  (shown in Fig.\ 1b of the main text) was defined by e-beam such that SAWs propagate in the [011] direction, parallel to a cleaved edge. After developing the resist (1:3 MIBK:IPA for $30 \units{seconds}$), the chip was exposed to oxygen plasma in a reactive ion etcher (RIE) to remove resist remaining in the trenches ($50 \units{sccm}$ at $50 \units{W}$ for 5 seconds). Next, $30\units{nm}$ of Al was deposited and promptly oxidized in $4 \units{Torr}$ of $\mathrm{O}_2$ without breaking vacuum. Mask lift-off was in acetone.  A second resist layer with $400/200\units{nm}$ MMA/PMMA was spin-coated, and two Doland bridges were e-beam written to be in capacitive contact the previous layer (red in Fig.\ 1 of the main text). After the same development and RIE steps, a double angle evaporation ($30\units{nm}$ and $60\units{nm}$) with a $4 \units{min}$ oxidation step in between formed the split-junction transmon, as shown in Fig.\ 1c.


\section{qubit-IDT connectivity}

\begin{figure*}[htb]
\begin{center}
\includegraphics[width=1\linewidth]{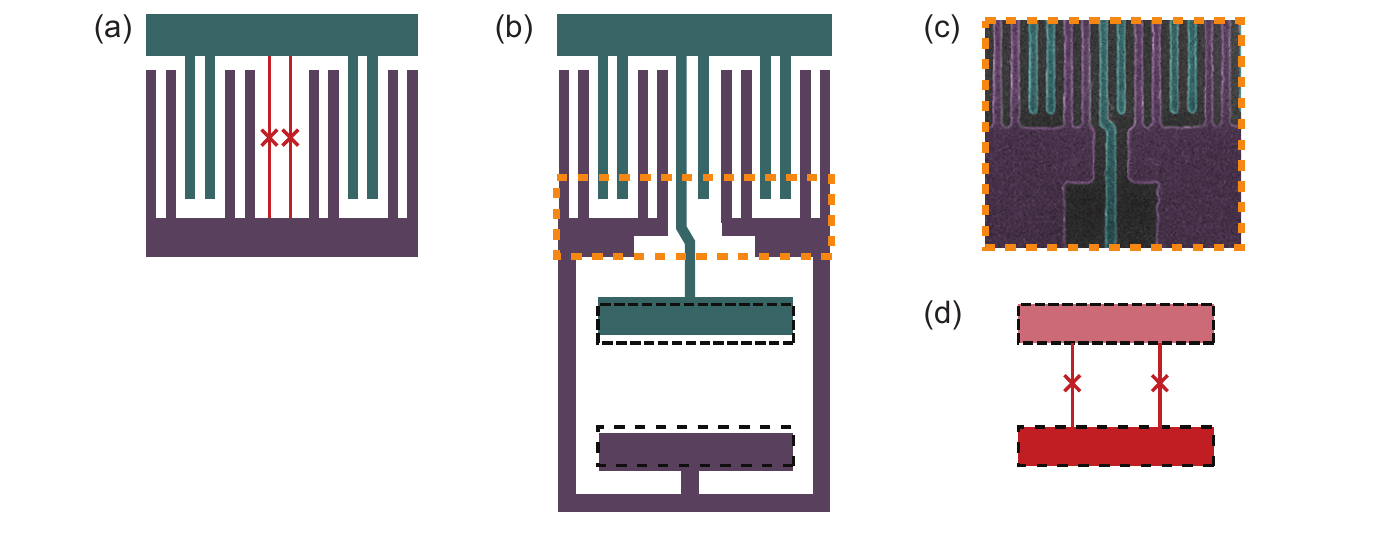}
\caption {
(a) Geometry of an acoustic qubit with a split-junction patterned inside the IDT.
(b) A schematic of an acoustic qubit with a finger pulled through the bottom electrode to impose a voltage difference between the electrodes across the junction loop.
(c) An SEM image showing a qubit-IDT finger pulled through the bottom electrode. The image corresponds to the orange dashed region in (b).
(d) The geometry of the qubit layer patterned on top of the acoustic layer. The location is indicated by the black dotted lines in panel (b). 
}
\label{figS1}
\end{center}
\end{figure*}

Placing the split-junction inside the acoustic cavity (see Fig.\ \ref{figS1}a) breaks IDT symmetry and may introduce unwanted interactions between sound and the junction. To have the IDT capacitance shunt the junctions with minimal acoustic impact, the qubit was fabricated with the geometry shown in Fig.\ \ref{figS1}(b) (adapted from \cite{Gustafsson2014}). One finger connected to the top electrode (green) was pulled through the bottom (purple) electrode. 
Two large pads were deposited in the acoustic layer that formed bottom plates of parallel plate capacitors. Oxygen was bled into the evaporator to form a clean oxide layer on the aluminum. The junction layer (Fig.\ \ref{figS1}d) was deposited directly on top of the two pads, forming the upper plate of the capacitors. The pads are $275 \units{\mu m ^2}$, and have an impedance of $3 \units{\Omega}$ at $4 \units{GHz}$.

\section{Qubit Anharmonicity}

\begin{figure*}[htb]
\begin{center}
\includegraphics[width=0.5\linewidth]{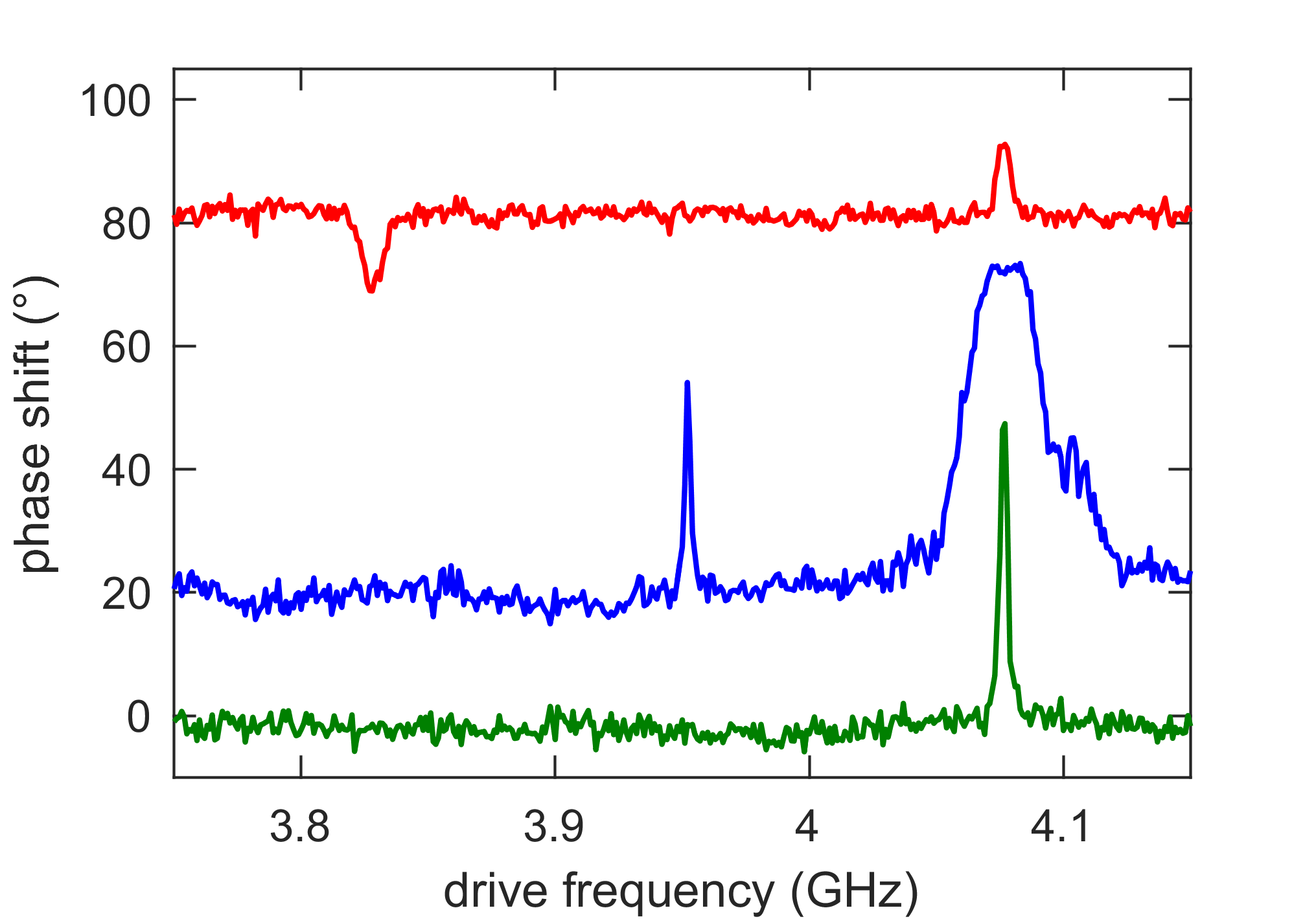}
\caption {Measurement of the transmon anharmonicity. One drive tone sweeps over the qubit frequency (x-axis), while a second tone reflects off the device near cavity mode $\omega_8$. The phase response of the second tone indicates the dispersive shift of the mode due to the qubit state. The green trace uses a low power qubit drive, blue is 20 dB higher in power, and red uses a third tone at $\omega_q$ to allow driving of the $|e\rangle - |f\rangle$ transition.
}
\label{figS2}
\end{center}
\end{figure*}

We measure the transmon anharmonicity $\alpha$ by varying the qubit drive power to observe other transition frequencies. The green trace shows low power qubit spectroscopy, where $\omega_\mathrm{q}$ is visible at $4.077 \units{GHz}$. Increasing the qubit drive power by 20 dB (blue) broadens the $|g\rangle - |e\rangle$ transition and the two photon transition $|g\rangle-|f\rangle$ appears at $3.952\units{GHz}$. This transition is expected to be $\alpha/2$ lower in frequency. In order to measure the $|e\rangle-|f\rangle$ transition, there must be population in the $|e\rangle$ state, and so a fixed-frequency, low power tone is applied at $\omega_\mathrm{q}$. Sweeping the drive tone, a dip is observed at the $|e\rangle-|f\rangle$ transition frequency, at $3.828 \units{GHz}$. From these two measurements, we determine $\alpha=249 \units{MHz}$. This is close to our predicted value of $E_c=200\units{MHz}$ based on an estimate of our IDT capacitance.


\section{Coupling Estimation}

\begin{figure*}[htb]
\begin{center}
\includegraphics[width=1\linewidth]{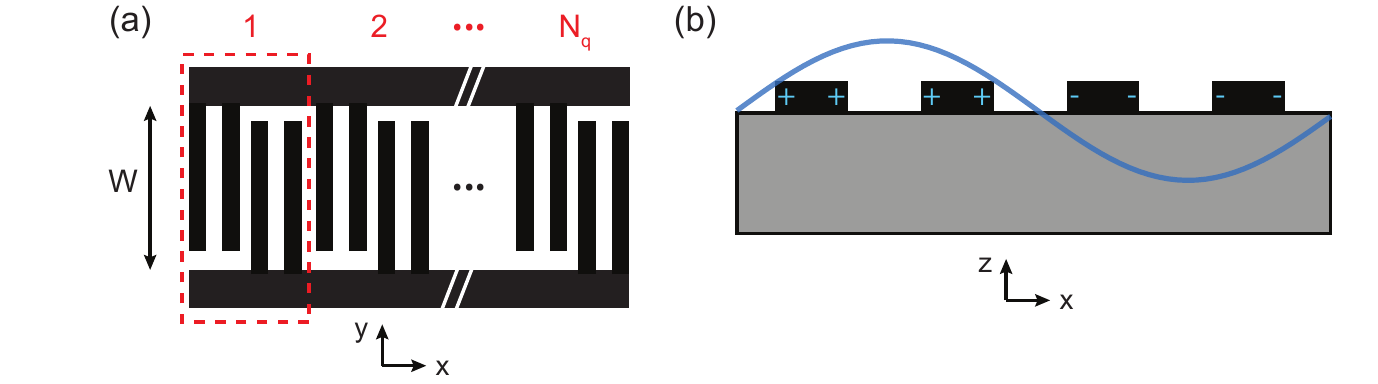}
\caption {Cross-section schematics of a split-finger qubit-IDT.
(a) The red dotted line highlights a unit cell of the IDT, corresponding the characteristic wavelength.
(b) One unit cell of a split-electrode IDT. The blue sine wave indicates the spatial profile of the SAW electric potential with wavelength $\lambda_c$. The SAW voltage interacts with the charges on the fingers, which are assumed to reside on the edge of the electrodes
}
\label{figS3}
\end{center}
\end{figure*}

The transmon coupling strength to the $m^{th}$ acoustic mode can be estimated by computing the electrical interaction energy between the two systems
\begin{equation}
\tag{S1}
g_m \hbar = \int d x^3 \rho_q V_m 
\end{equation}
where $\rho_q$ is the charge density of fluctuations on the qubit-IDT and $V_m$ is the zero point voltage fluctuations of the acoustic mode.
%
%
This integral can be greatly simplified by making several assumptions about the IDT and cavity geometries (see Fig.\ \ref{figS2}): (1) The IDT fingers are infinitely thin, and can be described by a delta function $\delta(z=0)$ in the $z$-direction. (2) The IDT and acoustic mode are uniform over the cavity width $W$ in the transverse $y$-direction, which can be written as a rectangle function $\Pi(y,y{-}W)$. (3) The charge fluctuations are localized to the edges of the IDT fingers, resulting in a sum of $\delta$-functions over the number of finger edges in the $x$-direction. (4) The acoustic modes are purely sinusoidal. With these assumptions
\begin{equation}
\tag{S2}
\rho_q(x,y,z) \approx \frac{Q_0}{W} \delta(z{=}0)\Pi(y,y{-}W)\frac{1}{8 N_q}\sum_i^{8 N_q}{\delta(x{=}x_i) p_i}
\end{equation}
\begin{equation}
\tag{S3}
V_k(z=0) \approx \Phi_0 \sin(k_m x) \Pi(y,y{-}W)
\end{equation}
where $Q_0$ is the magnitude of charge fluctuations across the qubit-IDT, $N_q$ is the number of finger periods in the qubit-IDT, $p_i$ is the IDT connectivity ($p_i=1$ for a finger connected to the top of the IDT, and $p_i=-1$ corresponds to a finger connected to the bottom), $\Phi_0$ are the zero-point voltage fluctuations of the acoustic mode at the surface, and $k_m$ is the wave number of mode $m$. Substituting Eq.\ (S2) and (S3) into Eq.\ (S1) simplifies the integral to a sum over the number of fingers edges
\begin{align*}
g_m\hbar &= \Phi_0 Q_0 \frac{1}{8 N_q}\sum_i^{8 N_q} \sin(k_m x_i) p_i\\
&= \Phi_0 Q_0 S_m
\end{align*}
%
where $S_m$ is an array factor that quantifies the spatial alignment between the IDT and the $m^{th}$ acoustic mode. Assuming periodic IDT finger locations, $x_i=x_0+s n$ for an IDT centered $x_0$, finger edge spacing $s$, $n\in \mathbb{Z}$, and chosen $p_i$ for the split finger geometry, this array factor becomes
\[
S_m\approx \frac{\sin(\pi/8)+\sin(3\pi/8)}{2} \sin\left( \frac{m\pi}{L} x_0\right) \text{sinc}\left(N_q \frac{\omega_a-\omega_m}{\omega_a}\right)
\]
where $\omega_a=2\pi v_s/(8 s)$ is the SAW angular frequency corresponding to the periodicity of the qubit-IDT. The numerical term $\big[\sin(\pi/8)+\sin(3\pi/8)\big]/2\approx 0.65$ arises from the split finger IDT geometry. In this device, the $\text{sinc}$ term is approximately 1 within the $50 \units{MHz}$ bandwidth of the acoustic modes. The qubit position $x_0\approx L/4$ implies
\begin{align*}
\sin\left(\frac{m\pi}{L}\left(\frac{L}{4}+\delta \right)\right)&=\sin\left(\frac{m \pi}{4}+\frac{m \pi \delta }{L}\right)\\
&\approx\sin\left( \frac{m\pi}{4}+\phi_q \right)
\end{align*}
where $\delta$ is the distance from the IDT center to $L/4$ and $\phi_q$ is a mode-independent phase offset. While $m \pi \delta$ might not be small compared to $L$, $\phi_\mathrm{q}$ can be considered a constant as long as it does not change dramatically over the range of modes, i.e. $\Delta_m \pi \delta \ll  L$, where $\Delta_m$ is the number of modes considered.

The zero-point voltage fluctuations $\Phi_0$ of the cavity are
\[
\Phi_0=\frac{e_{pz}}{\epsilon}\sqrt{\frac{\hbar}{2 D v_s A}}
\]
where $D$ is the density of the substrate, $A$ is the effective cavity area, $v_s$ is the speed of sound, $e_{pz}$ is the piezoelectric coefficient, and $\epsilon$ is the substrate permitivity. The magnitude of the charge fluctuations across the IDT is
\[
Q_0=2 e \beta \frac{1}{\sqrt{2}}\left(\frac{E_j}{8 E_c}\right)^{1/4}
\]
where $e$ is the electron charge, $E_J$ is the transmon Josephson energy, $E_C$ is the charging energy, and $\beta$ is a capacitive ratio $\beta \approx 1$ where $C_{IDT}$ is the capacitance of the IDT and $C_{\Sigma}$ is the sum of all capacitive elements that couple to the the junctions.
Therefore, the coupling has the form
\begin{align*}
g_m  &\approx 0.65 \frac{\Phi_0 Q_0}{\hbar} \sin\left(\frac{m \pi}{4}+\phi_q\right)\\
&\approx \frac{1}{\hbar}\frac{e_{pz}}{\epsilon}\sqrt{\frac{\hbar}{2 D v_s A}}\times e \left(\frac{E_j}{2 E_c}\right)^{1/4}\times 0.65 \sin{\left(\frac{m \pi}{4}+\phi_q\right)}\\
&\approx \sin\left(\frac{m \pi}{4}+\phi_q\right)\times2\pi\times 8.3 \units{MHz} 
\end{align*}
which is close to the measured value of $2\pi\times 6.48 \units{MHz}$. It is interesting to note that the number of fingers in the qubit-IDT is present only in decreasing $E_c$ and has little effect on the coupling strength as long as $\beta\approx1$. Decreasing the cavity area by a factor of 10, however, could lead to a 3-fold increase in coupling strength.


\section{Resonant Interaction Model}

To fit the reflection spectrum with the qubit crossing the acoustic modes (Fig.\ 2a of the main text), first the bare acoustic spectrum is measured with the qubit far detuned. This fixes the resonant frequencies $\omega_m$ and internal loss rates $\kappa_{m,\mathrm{in}}$ of all the modes measurable by the cavity-IDT. The external coupling rates $\kappa_{m,\mathrm{ex}}$ agree with predictions of how an IDT centered within the cavity should couple
\begin{align*}
\kappa_{m,\mathrm{ex}}&=\kappa_0\left[\sin\left(\frac{m \pi}{2} +\phi_c \right) \text{sinc}\left(\pi N_c \frac{f_m-f_c}{f_c}\right)\right]^2\\
&=\kappa_0 a_m^2
\end{align*}
where $f_c$ is the center frequency of the cavity-IDT, and $f_m$ is the mode frequency, $N_c$ is the number of finger periods in the cavity-IDT, $\phi_c$ accounts for a deviation in the cavity position from the cavity center, and we define $a_m$ to carry all $m$ dependence. Fitting the external coupling rates to the model gives $\kappa_0= 178.2 \units{kHz}$,  $\phi_c=\pi/4 -0.09$, and $f_0=4.253 \units{GHz}.$ The value of $\phi_c\approx\pi/4$ means the IDT couples almost equally to both even and odd modes. The external coupling rate is not affected by the sign of $a_m$, 
however when the qubit interacts with these modes causing them to hybridize, the relative sign between the various $a_m$ matters. 
Using the best fit values of the external coupling allows us to access the relative sign changes in $a_m$ needed when calculating the external coupling rates of acoustic mode superpositions.

The qubit frequency at each flux bias must also be input into the model. The transition frequency as a function of applied current is
\begin{equation}
\omega_{\mathrm{q}}=\omega_{\mathrm{max}}\sqrt{\left|\cos\left(\pi\frac{I-I_b}{ I_0} \right)\right|} .
\end{equation}
%
where $\omega_\mathrm{max}$ is the maximum qubit frequency, $I_0$ is the current in the coil needed to thread 1 flux quanta through the split junction, $I$ is the applied current, and $I_b$ is the current needed to offset the background flux. Qubit spectroscopy was performed (see Fig.\ \ref{figS5}) to fit these values, yielding $\omega_{\mathrm{max}}=2\pi\times 5.08 \units{GHz}$ and $I_0=1.0395 \units{\text{mA}}$. The offset $I_b$ changes a small but noticeable amount over the duration a day ($\approx 1 \units{\mu A}$), and so was left as a free parameter in the fit.

The eigenstates at each coil bias point can be found by diagonalizing the Hamiltonian of the coupled system
\[
 H = \begin{pmatrix} \omega_1  &  &  & & & g_1\\  &   \omega_2  &  & &&g_2\\  &  &   \omega_3  & &&g_3\\  &  &   &  \omega_4&&g_4\\  &  &   &  & \ddots  & \vdots \\ g_1  & g_2  & g_3  & g_4 & \cdots  &  \omega_{\mathrm{q}}\end{pmatrix} 
\label{SuppHamiltonian}
\]
where $g_m$ is the coupling strength between mode $m$ and the qubit. Although there appears to be many free parameters, the coupling strength dependence on mode number is sinusoidal
\[
g_m = g_0 \sin\left(\frac{m \pi}{4}+\phi_0\right)
\]
This reduces 11 longitudinal mode parameters down to 2. The strength of the qubit coupling to transverse modes relative to longitudinal modes can be inferred from measurements made with the cavity-IDT. As the cavity-IDT couples more strongly to the purely longitudinal modes than those with a non-zero transverse mode number, we also expect the qubit-IDT to couple similarly. For the cavity-IDT, the ratio of longitudinal coupling $a_m$ to transverse $a_{m,\mathrm{tr}}$ is $a_{m,\mathrm{tr}}/a_m\approx0.35$. We assume the same ratio in longitudinal to transverse coupling for the qubit-IDT so that $g_{m,\mathrm{tr}}=0.35 g_m$.

The eigenvalues $E_i$ of this matrix correspond to the hybridized resonant frequencies of the acoustic-transmon system. The eigenvectors $c_{ij}$ describe superpositions of acoustic modes and qubit that form the new eigenmodes. The new internal and external coupling rates $\kappa'_{m,\mathrm{in}}, \kappa'_{m,\mathrm{ex}}$ are calculated by taking
\begin{align*}
\kappa'_{m,\mathrm{ex}}&=\kappa_0 \left(\sum_{n=1}^{17} a_n c_{mn}\right)^2\\
\kappa'_{m,\mathrm{in}}&= \sum_{n=1}^{17} \kappa_n c_{mn}^2 + \gamma c_{m,18}^2.
\end{align*}
We then calculate the reflection from the multiple acoustic modes using the eigenvalues and new internal and external coupling rates. The possibility of interference in the external coupling due to the possible sign changes in both $a_n$ and $c_{mn}$ account for the prominent change in mode visibility at the crossing. This corresponds to the hybridized modes having a different spatial overlap with the cavity-IDT than the bare modes. The modes that disappear have destructive interference at the IDT fingers and thus have no overlap with the cavity-IDT, while the visible modes have constructive interference and thus stronger external coupling than the bare modes.




\section{dispersive limit qubit spectroscopy}

\begin{figure*}[htb]
\begin{center}
\includegraphics[width=0.5\linewidth]{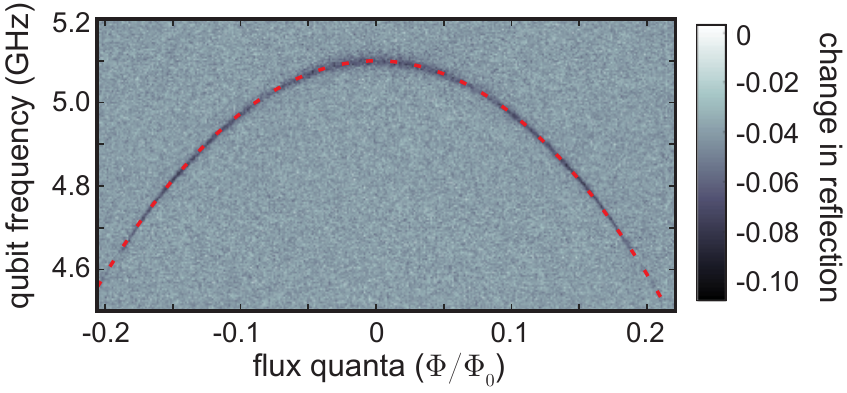}
\caption {
Tuning of the first qubit transition frequency ($\omega_{\mathrm{q}}$) with magnetic flux above the cavity resonances. The red dotted line shows the qubit frequency fit used in Fig.\ \ref{fig2}(a) of the main text with a small background field offset, indicating consistency between the on-resonance and dispersive regime measurements.
}
\label{figS5}
\end{center}
\end{figure*}

For each value of applied magnetic flux, we apply one tone near the resonant frequency of cavity mode 8 and monitor the magnitude of its reflection while sweeping a second tone through the qubit's resonance frequency. We attribute the change in cavity reflection with a qubit state dependent dispersive shift of the cavity frequency. In Fig.\ \ref{figS5}, we plot the change in cavity reflection versus qubit drive frequency, observing that the qubit's resonance tunes with flux as expected.


\section{Higher-order Dispersive Interaction}

To calculate the dispersive shift with 4 transmon levels and two-phonon processes, we numerically diagonalize the Hamiltonian describing a harmonic oscillator coupled to the transmon,
\begin{equation}
\tag{S4}
H=a^\dagger a \omega_8 +\sum_{i=0}^3 |i\rangle\langle i| E_i+\sum_{i=0}^2 \sqrt{i+1} g_8\left(|i\rangle\langle i+1| a^\dagger + |i+1\rangle\langle i| a\right)
\end{equation}
where $|i\rangle$ are the transmon levels with energy $E_i$, and $a$ ($a^\dagger$) is the annihilation (creation) operator for cavity mode 8. The effective coupling strength increases with $i$ due to the cooper pair number operator matrix element $\langle i|\hat{n}|i+1\rangle \propto \sqrt{i+1}$. The energy levels were assumed to follow $E_i=i \omega_\mathrm{q} + i (i-1)) \alpha/2$, where $\alpha$ is the anharmonicity of the qubit.

\begin{figure*}[htb]
\begin{center}
\includegraphics[width=0.5\linewidth]{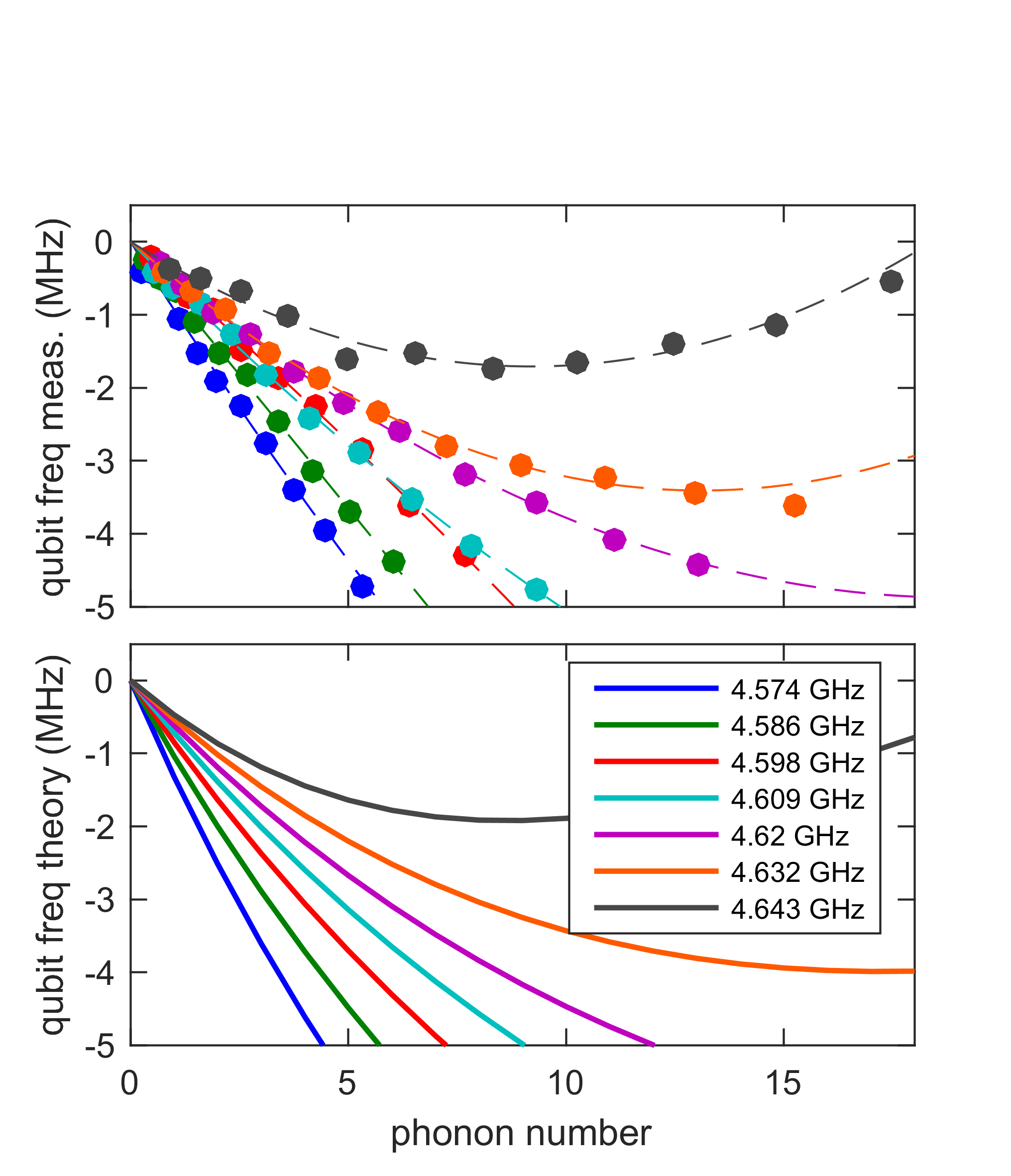}
\caption {
Measured and predicted quadratic stark shift (a) The measured qubit frequency shift (circles) versus phonon number for several qubit frequencies near the pole at $4.7 \units{GHz}$ or $\omega_8/2\pi+3\alpha /2$ with a quadratic fit (dotted lines). Color pattern for the qubit frequency matches the lower panel.
(b) The predicted qubit frequency shift is shown as solid lines. We see qualitative agreement between our data and theory predictions from numerical diagonalization of the model Hamiltonian.}
\label{figS4}
\end{center}
\end{figure*}
Writing this Hamiltonian in matrix form, the entries become block diagonal where the excitation number is conserved. In the $n^{th}$ excitation manifold, the block is
\[
 H_n = \begin{pmatrix} E_4 + (n-4) \omega_8  & 2 \sqrt{n-3} g_8  &  & & \\ 2 \sqrt{n-3} g_8 &   E_3 +(n-3) \omega_8 &  \sqrt{3(n-2)} g_8  & & &\\  &\sqrt{3(n-2)} g_8 & E_2+(n-2)\omega_8  &\sqrt{2(n-1)} g_8 &\\  &  &  \sqrt{2(n-1)} g_8 & E_1 + (n-1)\omega_8 &\sqrt{n} g_8&\\  &  &   & \sqrt{n} g_8  & E_0 +n \omega_8 \end{pmatrix} 
\]
This matrix was truncated at 50 phonons and numerically diagonalized. In the dispersive limit, the eigenvalues can be labelled effectively in the joint Fock basis between the acoustic mode and the transmon, where $E(i,j)$ has $i$ phonons in the oscillator and the transmon is in state $|j\rangle$. Then, the $|g\rangle-|e\rangle$ transition frequency with $i$ phonons is,
\[
\omega_{\mathrm{q}}(i)=E(i,1)-E(i,0)
\]
From this, $\chi$ can be written as
\[
2\chi(i) = \omega_\mathrm{q}(i+1)-\omega_\mathrm{q}(i)
\]
When pairs of energy levels are degenerate, $\chi$ diverges, and the dispersive approximation becomes invalid. The features that are most prominent are when $E(i,1)=E(i+1,0)$ and $E(i,2)=E(i+1,1)$. These are both first order as they only involve one exchange of excitation. However, two features present in our measurements occur when $E(i,3)=E(i+2,1)$ at $\omega_\mathrm{q}=\omega_8+3 E_c/2$ and when $E(i,2))=E(i+2,0)$ at $\omega_\mathrm{q}=\omega_8+E_c$. These are two-phonon processes and scale as $n^2 g^4/\Delta^3$ and have appreciable effects with small detuning and large phonon number.

Interestingly, the qubit frequency shift with phonon number becomes significantly quadratic near these degeneracy points. The predicted quadratic dependence of the qubit frequency shift qualitatively matches the predictions from diagonalizing the Hamiltonian (see Fig.\ \ref{figS4}).


\section{phonon emission rate}

The relationship between the IDT geometry and the phonon emission rate of the qubit
can be understood by considering the acoustic response of the IDT in the time domain. An impulse applied to the IDT generates a SAW wave with spatial properties that match the IDT. The resulting wave can be approximated as a sinusoid with wavelength $\lambda_c$ over a length $N_q \lambda _c$, where $N_q$ is the number of finger periods in the qubit-IDT. The wave propagates at $v_s$, creating a time-domain pulse with frequency $f_c$ and square envelope of duration $N_q / f_c$. The Fourier transform of this pulse gives the frequency dependence of the phonon emission rate of $\sin(X)^2/X^2$ with $X=N_q \pi (f-f_c)/f_c$. Therefore, an IDT has frequencies where SAW emission is prohibited, and an IDT with more fingers periods creates a sharper frequency response with more closely spaced nulls.

To quantify the overall scaling of this emission rate, we consider the qubit as a parallel RLC circuit with conductance due to SAW radiation,
\[
G(f) = G_{0} \left(\frac{\sin(X)}{X}\right)^2
\]
\[
G_{0}=2\pi\times1.3 K^2 N_q^2 f_{\mathrm{c}} W C_s
\]
where $N_q$ is the number of finger periods in the qubit-IDT, $C_s$ is the capacitance per unit finger length, $f_c$ is the center frequency of the cavity-IDT, $K^2$ is the piezoelectric coupling coefficient,  and $W$ is the width of the cavity. This gives a loss rate for the qubit equivalent RLC circuit
\begin{align*}
\Gamma_{\mathrm{SAW}}&=\omega \frac{G(f)}{2} \sqrt{\frac{L_j}{C_{\mathrm{IDT}}}} \\
&=\frac{2\pi\times 1.3 K^2 N_q f_{c}}{2 \sqrt{2}} \left(\frac{\sin(X)}{X}\right)^2\\
&= \Gamma_{\mathrm{max}}\left(\frac{\sin(X)}{X}\right)^2
\end{align*}
where $L_j$ is the junction inductance, and $C_{\mathrm{IDT}}$ is the cavity-IDT capacitance. For our device, $N_q=24$, $f_{c}=4.253 \units{GHz}$, $K^2=0.07\%$, we calculate $\Gamma_{\mathrm{max}}/2\pi=32 \units{MHz}$ . Using the lowest measured qubit linewidth as a frequency independent loss rate of $1.1 \units{MHz}$, we find excellent agreement between the predicted emission rate and the measured qubit linewidth as a function of qubit frequency.

\end{document}